\documentclass[preprint,aps,12pt,preprintnumbers,eqsecnum,nofootinbib,superscriptaddress]{revtex4}

\usepackage{amssymb,amsmath}
\usepackage[mathscr]{euscript}
\usepackage{mathrsfs} 
\usepackage[usenames,dvipsnames]{color}
\usepackage{float, subfigure}
\usepackage{setspace}
\usepackage{graphicx}

\usepackage{simplewick} 
\usepackage{epstopdf}

\everymath{\displaystyle}
\renewcommand{\cal}{\mathcal}

\newcommand{\bfX}{{\mathbf X}}
\newcommand{\bfb}{{\mathbf b}}
\newcommand{\bfk}{{\mathbf k}}

\begin{document}
%
\title{\vspace*{0.5in} 
Stochastic Emergent Quantum Gravity
}
\author{Joshua Erlich}\email[]{erlich@physics.wm.edu}\affiliation{Department of Physics,
William \& Mary, P.O. Box 8795, Williamsburg, VA 23187-8795, USA
}

\date{July 18, 2018}
%
%

%
%

\begin{abstract}
We present a stochastic framework for emergent quantum gravity coupled to matter.
The Hamiltonian constraint in diffeomorphism-invariant theories demands the identification of a clock relative to which dynamics may be defined, and other degrees of freedom can play the role of rulers. However, a global system of clock and rulers is generally not available.
We provide evidence that stochasticity associated with critical points of clock and ruler fields can be related to the emergence of both a probabilistic description consistent with ordinary quantum theory, and gravitation described by general relativity at long distances. 
We propose a procedure for embedding any Lorentz-invariant field theory, including the Standard Model and its Lorentz-invariant extensions, in this framework.

\end{abstract}

\pacs{}
\maketitle


\section{Introduction}
Spacetime appears to play conflicting roles in the two great frameworks of modern physics, quantum field theory and general relativity.
In quantum field theory, spacetime provides a parametrization for dynamics. We ask about correlations between observables at different points in spacetime. 
In the Schr\"odinger representation, the Schr\"odinger equation describes evolution of the quantum state relative to a parameter that we call ``time.'' However, in general relativity spacetime is itself the object of the dynamics. With respect to what clock does time-evolution refer if the quantum state provides a probabilistic description of spacetime itself?  
Initial attempts at quantization of general relativity provide an uncomfortable response to this puzzle: In diffeomorphism-invariant theories such as general relativity, the Hamiltonian vanishes up to boundary terms and constraints, and in spacetimes without boundaries the Schr\"odinger equation is replaced with a constraint  schematically of the form ${\cal H}|\psi\rangle=0$. The Wheeler-DeWitt equation of canonical quantum gravity \cite{DeWitt:1967yk} is a constraint of this type. States are not dynamical with respect to any external time parameter, so the problem of reconciling different notions of time-evolution would appear to be moot.  However, the timeless description of the quantum state only underscores the basic puzzle of how conventional notions of time might arise from a background-independent 
formulation of quantum gravity.

The {\em a priori} absence of dynamics in diffeomorphism-invariant quantum theories
suggests that an internal degree of freedom should be identified as a clock relative to which dynamics may be defined. Additional degrees of freedom might be identified as rulers, and the clock and rulers then provide a spacetime backdrop for the theory. Physical questions become relational: What is the correlation between Observable A when the clock and rulers have values $(T_A,\mathbf{X}_A)$, and Observable B when the clock and rulers have values  $(T_B,\mathbf{X}_B)$?   Such a Machian view of dynamics has been embraced as  a cornerstone of the search for a background-independent quantum theory of gravity by Barbour \cite{Barbour,Barbour1,Barbour2,Barbour:1994ri,Barbour:2000qg,Barbour:2010dp}, Page and Wootters  \cite{Page:1983uc},  Smolin \cite{Smolin:2005mq}, Rovelli \cite{Rovelli:1995fv} and others. 

But what makes for an appropriate identification of clock and rulers? It seems inevitable that any notion of space and time derived from quantum observables will inherit some form of quantum fuzziness. 
This may be avoided if the clock and rulers are classical or gauge-fixed, taking definite values, while the remainder of the system is quantum mechanical \cite{Peres:1997hz}. But even then, the mapping of coordinates to fields need not be invertible, and  a  globally defined  system of  clock and rulers is generally not available. This difficulty is referred to as the global problem of time  \cite{Kuchar:1991qf,Torre:1992rg,Isham:1992ms,Dittrich:2015vfa}.

The generic absence of a good system of clock and rulers hints that quantum mechanics and general relativity are not to be reconciled as fundamental aspects of the description of nature.
It is possible that both quantum theory and general relativity derive from a more primitive microscopic framework out of which the smooth, continuous dynamics of our experience emerges as an approximation.
This is not a new idea \cite{Smolin-QuantumConcepts}, although to the author's knowledge there is as yet no straightforward realization of a framework of emergent relativistic quantum field theory and general relativity. 

We claim that if  quantum field theory emerges from a more primitive framework that would become manifest at some short distance scale, then with little more than the presumption of diffeomorphism invariance in the fundamental framework, gravitation  is an emergent interaction at long distances. This claim may be considered an extension of Sakharov's induced gravity paradigm \cite{Sakharov:1967pk}, with the fundamental short distance scale playing the role of the regulator responsible for the induced gravitational interaction, which is quantized here rather than semiclassical as in the original induced gravity models. Diffeomorphism invariance, rather than creating a conflict between quantum theory and general relativity, instead becomes an integral component in the origin of them both.

As an illustration of this possibility, we propose that quantum theory and gravitation emerge from a stochastic dynamics of fields, including those fields that serve as clock and rulers.  
There are certain similarities between the  framework that we  develop here and the stochastic approach to quantum gravity proposed by Markopoulou and Smolin  \cite{Markopoulou:2003ps}. However, contrary to their construction based on graphs with discrete nodes, in the present approach there is no fundamental discreteness other than that of a stochastic evolution of fields, and physics is classical and nongravitational at ultra-short distances.  Furthermore, the present framework suggests a natural procedure for incorporating any Lorentz-invariant quantum field theory, although the description of  Grassmann (fermionic) fields \cite{Damgaard:1983tq}   and gauge fields in a stochastic setting  require further clarification.  We  note that the present framework is not, at least directly, related to stochastic gravity \cite{Moffat:1996fu,Hu:2008rga},  in which a stochastic evolution of the spacetime metric is postulated from the outset.

As a model for this picture, we are motivated by some recent analyses of induced emergent gravity models in which scalar fields were identified as a system of clock and rulers. In those models,  quantization was shown to give rise to scattering amplitudes that include long-range gravitational interactions consistent with general relativity \cite{Carone:2016tup,Chaurasia:2017ufl}, including gravitational self-interactions \cite{Carone:2017mdw}.  Similar models have also been studied from the perspective of the gravitational effective action \cite{Akama:1978pg,Akama:2013tua}, with a similar conclusion. 

In the analyses of scattering amplitudes in those models,  only configurations in which the scalar clock and ruler fields had profiles monotonic in a particular sense with respect to the spacetime coordinates were included in  computations. In other words, those fields were forced to act as a good system of clock and rulers. Other field profiles were argued to be nonperturbative in that approach, and their contribution to functional integrals determining scattering amplitudes was postponed to later consideration. Here we suggest that the additional field configurations, which would appear to make the choice of clock and ruler fields ill considered, do not contribute separately to functional integrals defining correlation functions. Instead, we propose that those configurations are related to a multivaluedness of physical degrees of freedom over spacetime that is associated with the emergence of the quantum-mechanical description in the first place.

We summarize our proposal for stochastic emergent quantum gravity by the following constructive procedure, which begins with some Lorentz-invariant  {\em classical} field theory:
\begin{enumerate}
\item Add to a classical $d$-dimensional Lorentz-invariant field theory $d$ massless and uncoupled scalar fields that we refer to as ``clock and ruler fields.'' The clock field has the wrong-sign kinetic term.
\item Embed the theory covariantly in an auxiliary spacetime.
\item Specify the auxiliary spacetime metric, or the vielbein up to local Lorentz transformations, as functions of the other fields by a constraint of vanishing energy-momentum tensor. An example of this step is the transition from the Polyakov action for the relativistic bosonic string to the Nambu-Goto action. 
\item The resulting theory is diffeomorphism invariant and background independent, and there are solutions to the equations of motion for arbitrary profiles of the clock and ruler fields. Assume a stochastic evolution of fields, including the clock and ruler fields. The typical spacing between stochastic impulses is ultimately related to the strength of the emergent gravitational interaction. 
The stochastic description is
modeled by the dynamics of Nelson's stochastic mechanics \cite{Nelson:1966sp,Nelson:1983fp} and its field-theory generalization by Guerra and Ruggiero \cite{Guerra:1973ck}. 
\end{enumerate}

The probabilistic description of the stochastic theory resembles the standard quantization of the original classical theory at distances and times much longer than the typical spacing between critical points. The discreteness of critical points provides a physical ultraviolet regulator for the emergent quantum field theory, and gravitation emerges at long distances as an artifact of the short-distance regulator.

\section{Why scalar clock and ruler fields?} \label{sec:ClockRulers}
Familiar examples of theories with scalar clock and/or ruler fields include the Nambu-Goto theory of the relativistic string, and the  covariant formulation of the free massive relativistic particle. We can illustrate the basic ideas in our construction by consideration of these theories, so we begin with discussion of the relativistic particle. The action is proportional to the
proper time elapsed on the particle's worldline, and  is given in a non-covariant form by,
\begin{equation}
\label{eq:Sparticle}
S_{particle}=-m\int dt\,\sqrt{1-\dot{{\mathbf X}}^2},
\end{equation}
where we use the dot notation $\dot{{\mathbf X}}\equiv d{\mathbf X}/dt$. The Euler-Lagrange equations for the particle position ${\mathbf X}$ are, \begin{equation}
\frac{d}{dt}\left(\frac{m\dot{\mathbf X}}{\sqrt{1-\dot{{\mathbf X}}^2}}\right)=0, \end{equation}
and solutions are of the form ${\mathbf X}(t)={\mathbf v}t+{\mathbf x_0}$ for constants ${\mathbf v}$ and ${\mathbf x_0}$.
This description is not manifestly Lorentz covariant, as the position degrees of freedom ${\mathbf X}$ play a different role than the time parameter $t$. 

In order to formulate the theory covariantly we  introduce a clock  $T(t)$, and write 
\begin{equation}
\label{eq:Sparticle-cov}
S_{particle}=-m\int dt\,\sqrt{\dot{T}^2-\dot{{\mathbf X}}^2}.
\end{equation}
Lorentz invariance of the action is now manifest, and acts as a global symmetry on the degrees of freedom $T$ and ${\mathbf X}$  (with speed of light $c=1$). The introduction of the clock $T(t)$ has not increased the number of degrees of freedom of the theory because in addition to the Lorentz symmetry, Eq.~(\ref{eq:Sparticle-cov}) is  invariant under arbitrary diffeomorphisms that take $T(t)\rightarrow T(t'(t))$ and ${\mathbf X}(t)\rightarrow {\mathbf X}(t'(t))$. Consequently, $T(t)$ can be considered to be a gauge degree of freedom. The equations of motion are now,
\begin{eqnarray}
\frac{d}{dt}\left(\frac{-m\dot{T}}{\sqrt{\dot{T}^2-\dot{{\mathbf X}}^2}}\right)&=&0, \label{eq:TEOM}  \\
\frac{d}{dt}\left(\frac{m\dot{\mathbf X}}{\sqrt{\dot{T}^2-\dot{{\mathbf X}}^2}}\right)&=&0. \label{eq:TXEOM} \end{eqnarray}
There are solutions  of the form ${\mathbf X}[T(t)]={\mathbf v}\,T(t)+{\mathbf x_0}$  for any 
$T(t)$ and constants ${\mathbf v}$ and ${\mathbf x_0}$. 

 Comparing the solutions of the equations of motion in the covariant and non-covariant descriptions, it should be evident that  $T(t)$ plays the role of a clock and provides an interpretation for time in the covariant formulation.  The parameter $t$ plays little role in the interpretation of the dynamics of the covariantly described motion; dynamics is interpreted in terms of the relation between ${\mathbf X}$ and clock time $T$. As long as $T(t)$ is monotonic it can be transformed to the static gauge $T(t)=t$ by a diffeomorphism.  In the static gauge we are left with the original, non-covariant, description of the theory. 
 
 While this example is  quite basic, it succeeds in the Machian 
 objective of identifying inertial frames based on the state of the universe rather than {\em a priori}. Here the relevant aspect of the state of the universe is the configuration of the clock  $T(t)$. A completely relational theory should also include rulers that together with the clock  provide a frame relative to which dynamics may be described.
   
With $T(t)$ treated as a dynamical degree of freedom, the Hamiltonian vanishes in the covariant description. However, there is a primary constraint (a constraint independent of the equations of motion) among the canonical momenta that defines the dynamics, namely $p_T^2-{\mathbf p}_{\mathbf X}^2=m^2$, with canonical momenta $p_T$ and ${\mathbf p}_{\mathbf X}$ defined from the action Eq.~(\ref{eq:Sparticle-cov}) in the conventional manner, so that:
\begin{eqnarray}
p_T=\frac{-m\dot{T}}{\sqrt{\dot{T}^2-\dot{{\mathbf X}}^2}}, \ \ \  {\mathbf p}_{\mathbf X}=\frac{m\dot{\bfX}}{\sqrt{\dot{T}^2-\dot{{\mathbf X}}^2}}. \end{eqnarray}
The primary constraint is preserved by the equations of motion. If we tentatively define the quantum theory by functional integral quantization in configuration space, but restricted to include only those profiles of $T(t)$ that can be transformed to static gauge by a coordinate transformation, then the resulting theory is equivalent to quantization of the non-covariant formulation of the theory. 
The Fadeev-Popov procedure allows $T(t)$ to be put in static gauge without the introduction of ghosts. Note that the condition $\dot{T}\neq0$ is invariant under nonsingular coordinate transformations, so that even with the restriction to monotonic $T(t)$  the description of the theory is generally covariant.

There are other approaches to quantization of constrained systems of this type. In Dirac's formalism, the constraint is added to the Hamiltonian multiplying a Lagrange multiplier, and a canonical quantization procedure is then possible.  Alternatively, with the constraint added to the Hamiltonian, the quantum theory can be defined by functional integration in phase space \cite{Halliwell:1988wc}. We generally focus on configuration-space descriptions because they more cleanly illustrate the Machian nature of these systems.

The discussion of the free particle  generalizes to theories with covariant interactions. For example, the relativistic particle coupled to a background electromagnetic field $A_M(T,\bfX)$ is described by the diffeomorphism-invariant action
\begin{equation}
S_{particle}=\int dt\,\left[-m\sqrt{\dot{T}^2-\dot{{\mathbf X}}^2}+q\dot{X}^M A_M(T,\bfX)\right],
\end{equation}
where we use the notation $X^0\equiv T$, $X^i\equiv\bfX^i$, $i\in\{1,\dots,d-1\}$. 
In terms of $X^M[T(t)]$ the Euler-Lagrange equations for this system can be written, \begin{eqnarray}
\dot{T}\left(\frac{d}{dT}\left[\frac{X^{M\prime}(T)}{\sqrt{1-\bfX'(T)^2}}\right]-qF^M_{\ N}\,X^{N\prime}(T)\right)=0 ,
\end{eqnarray}
where $F_{MN}\equiv\partial_MA_N(X)-\partial_NA_M(X)$ and indices are raised and lowered with the Minkowski metric $\eta^{MN}$. As in the free theory, there are solutions for any $T(t)$, and $T(t)$ plays the role of a clock in the  dynamics. The background electromagnetic field breaks Lorentz invariance, but not the diffeomorphism invariance of the description. One can check that the equation of motion for $T$ follows from the equations of motion for $\bfX$, which is consistent with the interpretation of $X^0$ as a gauge degree of freedom.

To summarize this discussion, the introduction of a clock  $T(t)$ in this (0+1)-dimensional field theory at the same time makes global Lorentz invariance manifest when it is a symmetry of the theory, and gives rise to a diffeomorphism-invariant description of the theory. The clock  does not add a physical degree of freedom to the non-covariant description of the theory; the new degree of freedom is eliminated by the diffeomorphism invariance of the covariant formulation. A subtlety in the covariant formulation is the possibility of branch points in the Lagrangian and in the canonical momenta if $\dot{T}=0$ somewhere along the trajectory of the particle. The  presence of such branch points will be relevant in what follows.

String theory, as a background-independent theory on the worldsheet (but not the target spacetime), provides another useful point of comparison to these ideas. The Nambu-Goto action for the bosonic string in $d$ spacetime dimensions is proportional to the area of the string worldsheet parametrized by $X^M(\tau,\sigma)$, $M\in\{0,\dots,d-1\}$:
 \begin{equation}
S_{{\rm string}}=-T_0\int d^2x \sqrt{\det\left(\eta_{MN}\frac{\partial X^M}{\partial x^\mu}\frac{\partial X^N}{\partial x^\nu}\right)},
\label{eq:StringAction}
\end{equation}
where  $x^0\equiv \tau$, $x^1\equiv \sigma$, and $\eta_{MN}$ is the Minkowski metric in mostly minus signature.
The action is invariant under diffeomorphisms $X^M(x)\rightarrow X^M[x'(x)]$ for any invertible coordinate transformation $x\rightarrow x^{\prime}(x)$. In order to analyze the dynamics of the string  it is convenient to choose a static gauge for which $X^0(\tau,\sigma)=\tau$. The coordinate $\tau$ is taken to range over the real line. 

The static gauge choice is possible because we {\em assume} that the action describes the worldsheet area of a string. This is a restriction on the class of  configurations of the fields $X^M(x)$. For example, if in some parametrization $X^0(x)=$~constant, then no reparametrization of coordinates $\tau$ and $\sigma$ will make the scalar field $X^0$ anything other than its constant value, and there would be no static gauge. A similar conclusion follows if $X^0$ has critical points at which derivatives with respect to the worldsheet coordinates vanish. At the critical points the Lagrangian vanishes, and from the target-space perspective there is no timelike tangent vector at such points \cite{Zwiebach:2004tj}. From the perspective of the (1+1)-dimensional worldsheet field theory, restriction to static gauge is a physical restriction on the space of allowed field configurations.  We note that the  possibility of topology-changing and metric-changing configurations in general relativity, strings and branes, which suffer similar questions of interpretation, have long been of interest~\cite{Geroch:1967fs,Sakharov:1984ir,Hayward:1992zp,Rey:1998yx,Gibbons:2004dz,Dijkgraaf:2016lym}.

\section{Jittery clocks and stochastic mechanics} \label{sec:StochasticMechanics}
In the covariant description of the relativistic particle, if the clock  profile $T(t)$ is non-monotonic then its interpretation as an identifier of time would appear to be problematic. There may be different coordinate values $t_i$ that correspond to the same clock time $T$, so physical degrees of freedom would visit the same clock time more than once while scanning coordinate time $t$. In the generalization to higher-dimensional field theories, we assume that there are both clock and ruler fields relative to which spacetime dynamics is defined. In that case, the global problem of spacetime is that generically  there are  critical points $x_{\rm crit}$  such that $\det\left(\partial X^I/\partial x^\mu\right)|_{x_{\rm crit}}=0$, in which case  the mapping from coordinates $x^\mu$ to fields $X^I(x^\mu)$ is in general not invertible, and 
the remaining fields in the theory could be multivalued if considered  functions of the values of the clock and ruler fields. 

As  a better system of clock and rulers does not appear to be available, 
we choose to embrace the possibility that nature may have granted us only jittery clocks and folded rulers, and follow this rabbit hole where it leads. 
We will find that critical points of clock and ruler fields fit naturally into a stochastic description of quantum fields.
We illustrate how stochasticity associated with critical points of clock and ruler fields  can be related to the uncertainties of quantum theory by considering again the covariant description of the  relativistic particle.
As discussed earlier, due to diffeomorphism invariance the equations of motion for $T$ and $\bfX$ are not independent, and there are solutions with arbitrary $T(t)$.

At critical times when $\dot{T}=0$ the progression of clock time ceases, and in general changes direction with respect to coordinate time $t$. At these critical times, the equations of motion force $\dot{\bfX}$ to zero, and the square root in the relativistic Lagrangian reaches a branch point. 
One possibility is that the solution $\bfX(T)$ remains 
single-valued at all clock times, even when passing through the branch points. This is what is suggested by the classical equations of motion, which demand continuity of the canonical momenta even through the branch points and determines the choice of branch upon passing through such points.   
In this scenario classical physics evolves both forward and backward in clock time $T$, but simply backtracks while moving backward in clock time, and then repeats the same motion while moving forward once again. The global problem of time is resolved in this classical setting because the dynamics requires that physical degrees of freedom depend only on the clock time, even as the hands of our fundamental clock fluctuate both forward and backward.  This is a curious situation, but is not particularly interesting except perhaps through a philosophical lens. 

Instead of this repetitive forward-backward motion, we will postulate that the canonical momenta $\mathbf{p}_\bfX$ and $\mathbf{p}_T$ receive stochastic impulses, which may themselves be responsible for critical points of the clock $T(t)$.   We do not  provide a mechanism for the impulses, which we  take to be an irreducible component of this framework. The stochastic impulses  for each degree of freedom may  occur at independent times unrelated to the clock profile,  but for simplicity of presentation we imagine that the impulses are imparted to all of the degrees of freedom precisely at the critical times. These random impulses are responsible for an inability to predict with certainty the trajectory of the particle, so our object of interest will be the probability distribution over the space of possible trajectories. 
By associating the impulses with the critical points of the clock  $T(t)$, or the clock and ruler fields in the field-theory context, the role of the critical points is to connect different solutions of the classical equations of motion, acting as  a localized  instanton of sorts.

\begin{figure}
\centering
\hfil
\subfigure[
]
{
\includegraphics[scale=.3]{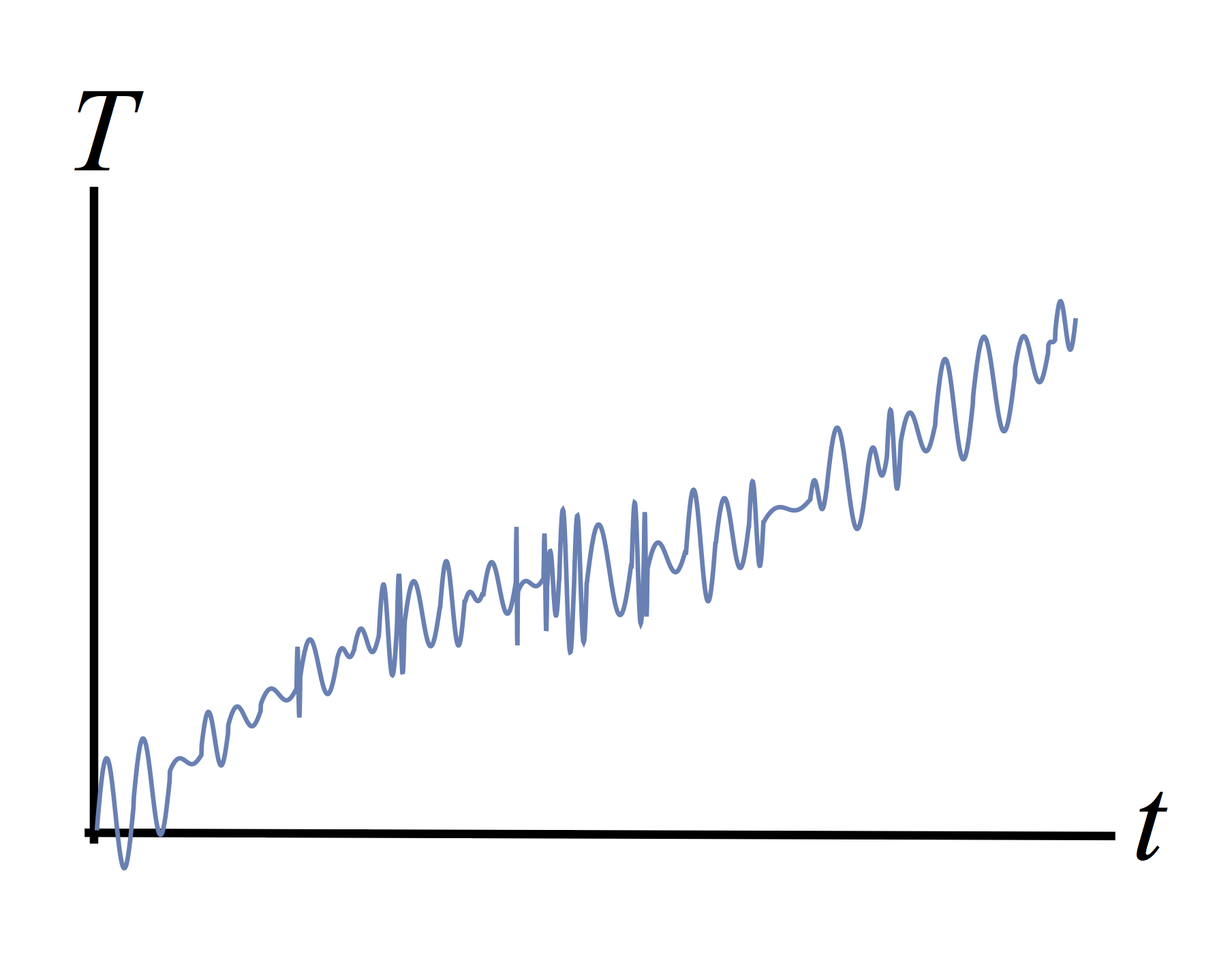}
\label{fig:Tt}
}
\hfil
\subfigure[
]
{
\includegraphics[scale=0.3]{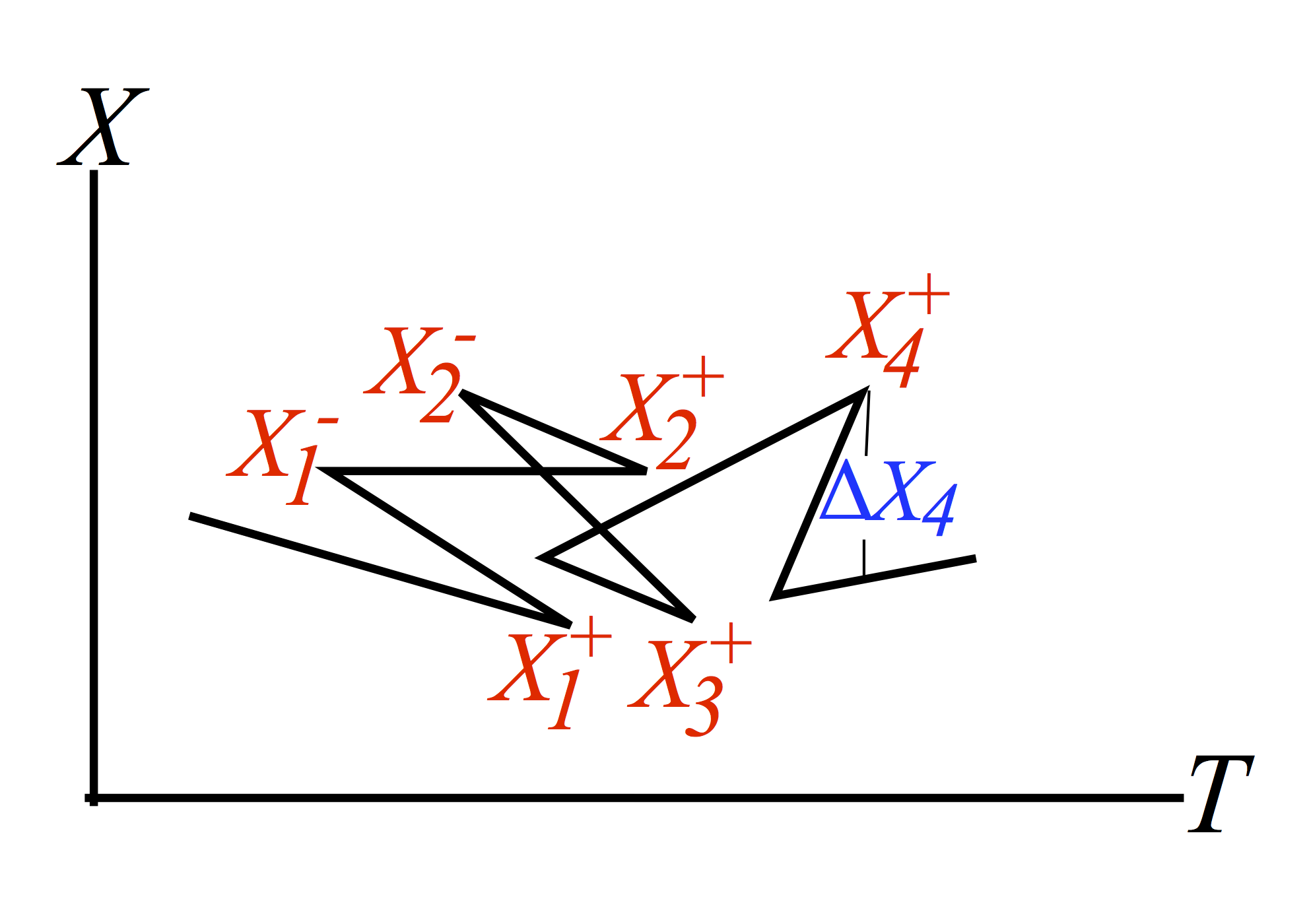}
\label{fig:XT}
}
\hfil
\subfigure[
]
{
\includegraphics[scale=0.3]{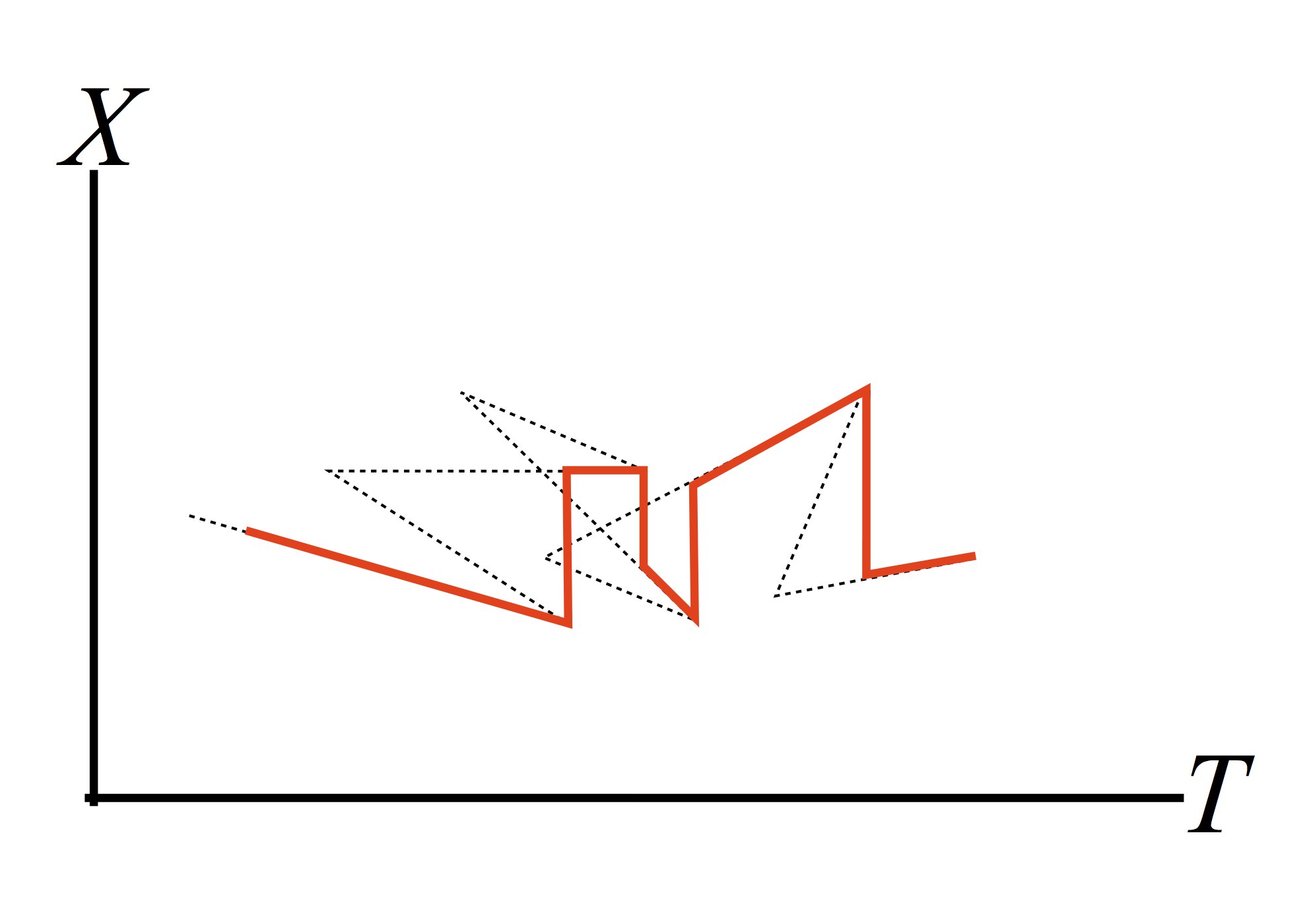}
\label{fig:XT2}
}
\hfil
\caption{(a) Random oscillatory evolution of clock time. The oscillations need not be regularly spaced in either $t$ or $T$, and can be assumed to take the form of a smoothed ({\em i.e.} differentiable) stochastic process with drift. (b) A forward-backward trajectory with position $\bfX_i^\pm$ at critical times $T_i^{\pm}$ labeled. The $\pm$ superscript refers to whether the critical point is at arrived at along the trajectory from the direction of increasing $T$ $(+)$ or decreasing $T$ $(-)$. The effective position jump $\Delta \bfX_4$ is indicated. There are effective jumps  corresponding to first and next arrival  at each  critical clock time $T_i^+$. (c) The same trajectory described by forward motion in clock time and discontinuous jumps in position.}
\label{fig:stochastic}
\end{figure}

We  still consider the particle position $\bfX$ to be dependent on clock time $T$, but as a consequence of the presumed impulses $\bfX(T)$ need not precisely retrace its steps as $T(t)$ regresses and then progresses forward again, and $\bfX(T)$ may now be multivalued as in Fig.~\ref{fig:XT}. One could instead imagine a stochastic evolution consistent with $\bfX$ depending only on the value of $T$, but one of our conclusions will be that the multivalued nature of physical degrees of freedom fits nicely into the framework of stochastic mechanics. Quantization in this scheme will refer to the dynamics imposed on the stochastic trajectories, and in a coarse-grained approximation we will describe what choices are necessary in order to reproduce the rules of standard quantum mechanics.

This stage of the discussion has not paid consideration to Lorentz invariance, as the role of the clock  has been distinguished from that of the position degrees of freedom.   As the present discussion is being framed in terms of particle mechanics  for simplicity of illustration, we do not pursue a  Lorentz-invariant framework at this stage. Lorentz invariance is more natural (but still at risk due to the discreteness of the impulses) in the higher-dimensional field-theory generalization of this picture because of the addition of ruler fields that play a similar role to the clock field. 
For the present discussion we will just assume that there is some non-dissipative distribution of impulses  at the critical points. 
The presumed absence of dissipation 
implies that the fluctuation-dissipation theorem, which would relate the stochastic diffusion constant to a frictional interaction strength, does not apply. The absence of dissipation is crucial here in that it allows for a notion of time-reversibility of the stochastic dynamics \cite{Nelson:1983fp}. 

We assume that there are many critical points  
between intervals of $T$ over which classical forces would cause significant change in motion. 
The particle explores a region in the space of trajectories by backtracking repeatedly in clock time and sampling different velocities $\dot\bfX(T)$ from nearby initial positions, before ultimately progressing forward in clock time and traveling an observably significant distance. There is a similarity between this sampling of paths and the path-integral approach to quantum mechanics, though they are conceptually distinct pictures. 

We can define an effective trajectory that moves only forward in clock time (Fig.~\ref{fig:XT2}), but because of the forward-backward trajectory in clock time and the stochastic behavior of $\bfX$, at each critical point there is in effect a random jump in position ({\em e.g.} $\Delta \bfX_4$ in Fig.~\ref{fig:XT}) in addition to the random change in velocity. At a microscopic level both the absence of dissipation and the effective jumps in position distinguish the process from classical Brownian diffusion, in which  each particle experiences effectively random impulses by interactions with molecules in a fluid, but the position of each particle is continuous. However, over mesoscopic scales between the typical difference between neighboring critical times and scales of experimental relevance, the effect of the microscopic jumps on the particle position are well modeled kinematically by a Brownian motion rather than a jump process.
The difference between Brownian diffusion and the present situation lies more in the dynamics of the diffusion than the kinematics.%

We constrain the stochastic distribution of impulses by comparison with Nelson's stochastic formulation of quantum mechanics \cite{Nelson:1966sp,Nelson:1983fp}. Forward and backward derivatives in the stochastic formalism now describe average rates of change forward and backward in clock time rather than coordinate time, but aside from minor changes in its interpretation the formalism of stochastic mechanics can be adopted in its entirety. 
The main result of Nelson's mechanics is that the probabilistic description of an ensemble of  stochastic trajectories satisfying Newton's second law $\mathbf{F}=m\mathbf{a}$, for an appropriately defined acceleration $\mathbf{a}$, is governed by the 
Schr\"odinger equation and the Born rule \cite{Nelson:1966sp}. Because the ideas underlying stochastic mechanics have maintained a level of interest among those interested in foundational issues, descriptions of stochastic mechanics appear in many places. Nice summaries with perspectives on challenges to the framework can be found in Ref.~\cite{Bacciagaluppi} and the early chapters of Ref.~\cite{delaPena}. 

We now explain how the stochastic system described here can realize Nelson's stochastic mechanics.  The essential observation is that the possibility of clock time running both forward and backward fits naturally into the framework of stochastic mechanics. Our goal is not to derive stochastic mechanics from a new framework, but to use stochastic mechanics to constrain the allowed probability distributions for the stochastic trajectories in the present setting.
The effective jumps in position due to the oscillations in clock time modify Nelson's description only slightly.

We decompose the effective forward motion in two components: the continuous components and the jumps $\Delta \bfX_i$.
We denote by $\bfb_i$ the (assumed) nearly constant velocity of the particle along the segment between $\bfX_{i-1}^-$ and $\bfX_i^+$ in the notation of Fig.~\ref{fig:stochastic}, {\em i.e.}, \begin{equation}
\bfb_i\equiv\frac{\bfX_i^+-\bfX_{i-1}^-}{T_i^+-T_{i-1}^-}. \end{equation}
We define the time-averaged velocity over the interval $\Delta T_{mn}$ between times $T_m^+$ and $T_n^+$  as the rate of change in particle position over that interval, with the discrete jumps removed:
\begin{eqnarray}
\langle\bfb_{mn}\rangle&\equiv&\frac{\bfX_n^+ - \bfX_m^+-\sum_{i=m}^{n-1}\Delta\bfX_i}{T_n^+-T_m^+}, \\
&=&\frac{\Delta\bfX_{mn}-\Delta\xi_{mn}}{\Delta T_{mn}}, \end{eqnarray}
where \begin{eqnarray}
\Delta\bfX_{mn}&\equiv\bfX_n^+ - \bfX_m^+, \\
\Delta\xi_{mn}&\equiv\sum_{i=m}^{n-1}\Delta\bfX_i. \end{eqnarray}
We will be interested in evolution of the particle position over times long compared to the typical difference between successive critical times. We therefore summarize the above relations by modeling the particle motion as a Markov process in a form that resembles a Langevin equation,
\begin{equation}
d\bfX=\langle\bfb\rangle\,dT+d\tilde{\boldsymbol{\xi}}. \label{eq:Langevin}
\end{equation}
Averaging over many jumps,  the stochastic variable $\tilde{\boldsymbol{\xi}}$ is modeled as an isotropic Brownian motion with Gaussian distribution, such that the ensemble average of $d\tilde{\boldsymbol{\xi}}(T)$ vanishes, and all correlations of products of $d\tilde{\boldsymbol{\xi}}$  vanish except the quadratic variance.

We note that $\langle\bfb\rangle$  in Eq.~(\ref{eq:Langevin}) is a {\em time-averaged} velocity over the interval $dT$ for a particular trajectory, not an {\em ensemble} average. Hence, $\langle\bfb\rangle$ is defined only along the trajectory of the particle, and Eq.~(\ref{eq:Langevin}) is not a stochastic differential equation that can be solved for general trajectories. To turn Eq.~(\ref{eq:Langevin}) into a stochastic differential equation we consider an ensemble of trajectories and replace $\langle\bfb\rangle$ by an ensemble average drift velocity $\bfb(\bfX,T)$. We assume that the displacements $d\boldsymbol{\xi}$ from the ensemble average remain Gaussian distributed, with
 \begin{eqnarray}
\mathbb{E}\left[d\xi^I(T)d\xi^J(T)\right]&=&2D\,\delta^{IJ}dT, \label{eq:quadratic} 
\end{eqnarray}
for some diffusion constant $D$.
The notation $\mathbb{E}$ refers to an ensemble average,  
which depends on the probability distribution for the stochastic trajectory, and $D$ is the diffusion constant for the process. The indices $I$, $J$ label components of the spatial vector $d\boldsymbol{\xi}$.
The averages of products of $d\boldsymbol{\xi}$ along non-overlapping infinitesimal time intervals vanish because of the independence of the impulses. 

The Langevin equation for this system is then 
\begin{equation}
d\bfX=\bfb(\bfX,T)\,dT+d\boldsymbol{\xi}. \label{eq:Langevin2}
\end{equation}
 The introduction of the ensemble here is used to parametrize our ignorance regarding the evolution of the forward-moving time-average velocity $\langle\bfb\rangle$ along the particle's trajectory. 
 Given an initial ensemble probability density $\rho(\bfX,T_0)$ and a form for $\bfb(\bfX,T)$, the Langevin equation predicts that the probability density $\rho(\bfX,T)$ for later times satisfies the corresponding Fokker-Planck equation, or Kolmogorov forward equation, \begin{equation}
\frac{\partial \rho}{\partial T}=-\nabla\cdot(\bfb\rho)+D\nabla^2\rho. \label{eq:Fokker-Planck}
\end{equation}
For readers unfamiliar with this hydrodynamic formalism, a short route from the Langevin equation to the Fokker-Planck equation is the following \cite{Jacobs}:  Consider some smooth function  $f(\bfX)$ that does not depend explicitly on $T$, taken along a  trajectory satisfying the Langevin equation Eq.~(\ref{eq:Langevin2}). One of the key results of the theory of Brownian motion is that the quadratic variation of the motion is proportional to the time elapsed $dT$, as in Eq.~(\ref{eq:quadratic}). As a result, we must be careful in Taylor expanding smooth functions, and we are led to an application of Ito's lemma \cite{Ito}, which  allows us to remove the expectation value in Eq.~(\ref{eq:quadratic}). (We ultimately only require expectation values to deduce the associated probability distribution, so this detail is not critical to what follows.) Using Eq.~(\ref{eq:Langevin2}) and expanding to ${\cal O}(dT,d\boldsymbol{\xi})$ we obtain,
\begin{eqnarray}
df(\bfX)&=&\nabla f \cdot d\bfX +\frac{1}{2}  \frac{\partial^2f}{\partial X^I\partial X^J} dX^I \,dX^J +\dots \\ \vspace{\baselineskip}
&=&\left(\nabla f\cdot \bfb +D\nabla^2f\right)dT + \nabla f\cdot d\boldsymbol{\xi} +\dots, \label{eq:Ito}
\end{eqnarray}
where the ellipses refer to terms of higher order in $dT$. Given an ensemble with probability distribution $\rho(\bfX,T)$, we can calculate the change in the average location of $\bfX$ over a time interval $dT$. Using the vanishing of $\mathbb{E}[g(\bfX,T)d\boldsymbol{\xi}_T]$ in the Ito convention, for which $d\boldsymbol{\xi}_T$ corresponds to the change in $\bfX$ over a differential time elapsed {\em after} time $T$, we have,
\begin{eqnarray}
\mathbb{E}[df(\bfX)]&=&d\mathbb{E}[f(\bfX)]\equiv \int d\bfX\,\rho(\bfX,T)\,f(\bfX)  \\
&=&dT \int d\bfX\,\rho(\bfX,T)\,\left(\nabla f\cdot \bfb +D\nabla^2f\right). \end{eqnarray}
In the first line we  used the definition of the expectation value and in the second line we also used Eq.~(\ref{eq:Ito}).
Written differently, \begin{eqnarray}
\frac{d}{dT}\int &&d\bfX\,\rho(\bfX,T)\,f(\bfX)=\int d\bfX\,f(\bfX)\,\frac{\partial \rho}{\partial T} \label{eq:FP1}\\
&&=\int d\bfX\,f(\bfX)\,\left(- \nabla\cdot(\bfb \rho) +D\nabla^2\rho\right), \label{eq:FP2} \end{eqnarray}
where we have integrated by parts to isolate $f(\bfX)$ in the integrands. But the function $f(\bfX)$ was arbitrary, so we are led to identify the integrands in Eqs.~(\ref{eq:FP1}) and (\ref{eq:FP2}), which gives the Fokker-Planck equation for the probability density function, Eq.~(\ref{eq:Fokker-Planck}).\footnote{To be more precise in our consideration of the effective position jumps, we may consider the  jumps as an added Poisson white noise in the Langevin equation rather than a contribution to the Gaussian white noise, with $\boldsymbol{\xi}_P(T)=\sum_{i=1}^{n(t)} \Delta \bfX_i \delta(T-T_i)$, where $n(t)$ is  Poisson-distributed and $\Delta \bfX_i$  is Gaussian distributed. The Fokker-Planck equation is modified in this situation \cite{Denisov:2008fx}, but can be shown to reduce to Eq.~(\ref{eq:Fokker-Planck}) for narrow distribution of $\Delta \bfX_i$. Though we assume an absence of large flights, the modification of the Fokker-Planck equation incorporating significant Poisson noise would likely lead to a dangerous modification of the emergent quantum theory.}

Because of the importance of time-reversibility of the dynamics in the quantum theory we follow Nelson in defining a Langevin equation for the same stochastic motion considered backwards in time, written in terms of $dT>0$ \cite{Nelson:1966sp,Nelson:1983fp}:
\begin{equation}
d\bfX=-\bfb_*(\bfX,T)\,(-dT)+d\boldsymbol{\xi}_*. \label{eq:Langevin3}
\end{equation}
In general $\bfb\neq\bfb_*$ and $d\boldsymbol{\xi}\neq d\boldsymbol{\xi}_*$, but $d\boldsymbol{\xi}_*$ is a Brownian process with the same diffusion constant $D$ that describes the distribution of $d\boldsymbol{\xi}$.
The corresponding backwards Fokker-Planck equation has the form \begin{equation}
\frac{\partial \rho}{\partial T}=-\nabla\cdot(\bfb_*\rho)-D\nabla^2\rho. \label{eq:Fokker-Planck2}
\end{equation}
Adding the forward and backward Fokker-Planck equations and defining \begin{equation}
\mathbf{v}\equiv\frac{\bfb+\bfb_*}{2} \end{equation}
gives a continuity equation for the probability distribution function: \begin{equation}
 \frac{\partial\rho}{\partial T}+\nabla\cdot\left(\mathbf{v}\rho\right)=0, \label{eq:continuity}
\end{equation}
while subtracting the two equations gives an equation that can be solved for $\bfb-\bfb_*$ up to a curl which is assumed to vanish:
\begin{equation}
\mathbf{u}\equiv \frac{\bfb-\bfb_*}{2}=D\frac{\nabla \rho}{\rho}. \label{eq:u}
\end{equation}

We still need a dynamical condition for acceptable stochastic functions $\bfb(\bfX,T)$ and $\bfb_*(\bfX,T)$, and hence for the allowed states in this system. This comes in the form of $\mathbf{F}=m\mathbf{a}$, but the classical forces only affect the non-stochastic component of the acceleration, so we need an appropriate interpretation of the acceleration $\mathbf{a}$. From the microscopic perspective, if not for the stochastic impulses, $\langle\bfb\rangle$ should satisfy the classical equation of motion itself. Hence, we expect the stochastic dynamics to be described in terms of ensemble averages for which the impulses average to zero. 

Nelson found the appropriate dynamical equation for $\bfb(\bfX,T)$ that leads to the Madelung equations \cite{Madelung}, which are a hydrodynamic representation of the Schr\"odinger equation \cite{Nelson:1966sp,Nelson:1983fp}. To that end, he defined average forward and backward stochastic derivatives as:
\begin{eqnarray}
D_+f(\bfX,T)&\equiv& \lim_{\delta T\downarrow 0}\mathbb{E}\left[\frac{f[\bfX(T+\delta T),T+\delta T]-f(\bfX(T),T)}{\delta T}\right] , \\
D_-f(\bfX,T)&\equiv& \lim_{\delta T\downarrow 0}\mathbb{E}\left[\frac{f(\bfX(T),T) - f[\bfX(T-\delta T),T-\delta T]}{\delta T}\right] .
\end{eqnarray}
The limit $\delta T\downarrow 0$ is taken in the model of the system as a continuous stochastic process, but because of the discreteness of the critical times we always have in mind $\delta T\gg\Delta T$, where $\Delta T$ is the typical difference between neighboring critical times.
Using Eq.~(\ref{eq:Ito}), and the analogy for differentials backwards in time, we have,
\begin{eqnarray}
D_+f(\bfX,T)&=&\frac{\partial f}{\partial T}+\bfb\cdot \nabla f+\frac{D}{2}\nabla^2 f, \\
D_-f(\bfX,T)&=&\frac{\partial f}{\partial T}+\bfb_*\cdot \nabla f-\frac{D}{2}\nabla^2 f.
\end{eqnarray}

The appropriate time-reversal-invariant dynamical condition for the stochastic system coupled to a conservative force $\mathbf{F}=-\nabla V(\bfX)$, assumed to vary only over distances much larger than $\delta T$, is the Nelson-Newton equation \cite{Nelson:1966sp,Nelson:1983fp}:
\begin{equation}
\frac{m}{2}\left(D_-D_+\bfX+D_+D_-\bfX\right) =\frac{m}{2}\left(D_-\bfb+D_+\bfb_*\right)=-\nabla V(\bfX) . \label{eq:a}
\end{equation}
The dynamical equation Eq.~(\ref{eq:a}), together with the forward and backward Fokker-Planck equations Eq.~(\ref{eq:Fokker-Planck}) and Eq.~(\ref{eq:Fokker-Planck2}), determine self-consistently the  drift velocities $\bfb(\bfX,T)$ and $\bfb_*(\bfX,T)$, and the probability density $\rho(\bfX,T)$ given an initial probability density.

Eq.~(\ref{eq:a}) indicates the appropriate definition for acceleration in stochastic mechanics: $\mathbf{a}=1/2\left(D_-D_+\bfX+D_+D_-\bfX\right)$.
There are other possibilities {\em a priori} for the definition of an acceleration in this setting, such as $1/2(D_+D_+\bfX+D_-D_-\bfX)$, but those expressions are not as local as the chosen definition, and in any case they generally lead to a nonlinear variation of the Schr\"odinger equation \cite{Davidson} or else a parabolic differential equation for the probability density without wavelike solutions \cite{delaPena}. Further motivation for Nelson's choice have been provided from several different perspectives, including a variational approach resembling an action principle for the stochastic process \cite{Guerra:1982fn}.

What Nelson showed  is that $\rho(\bfX,T)$ in the system defined by Eqs.~(\ref{eq:Langevin2}), (\ref{eq:Langevin3})  and (\ref{eq:a}) is described by the Madelung equations, which are equivalent to Eq.~(\ref{eq:continuity}) together with one additional equation and the assumption that $\mathbf{v}$ is a gradient,  \begin{equation}
\mathbf{v}\equiv 2D/\hbar\, \nabla S. \label{eq:v} \end{equation}
The additional Madelung equation resembles the Hamilton-Jacobi equation with an extra term: \begin{eqnarray}
\frac{\partial S}{\partial t}&+&\frac{1}{2m} (\nabla S)^2+V(\bfX)-\frac{\hbar^2}{2m}\frac{\nabla^2\sqrt{\rho}}{\sqrt{\rho}}=0. \label{eq:Madelung} \end{eqnarray}
The last term in Eq.~(\ref{eq:Madelung}) is known as the quantum potential.
With the definition $\psi\equiv\sqrt{\rho}\,e^{iS/\hbar}$, the Madelung equations were originally formulated as a hydrodynamic description of the real and imaginary parts of the Schr\"odinger equation, \begin{equation}
i\hbar\frac{\partial \psi}{\partial t}=-\frac{\hbar^2}{2m}\nabla^2\psi+V(\bfX)\psi. \end{equation}

The Born rule, $\rho=|\psi|^2$ is satisfied by construction in stochastic mechanics, and is not an additional assumption. Different states of the system correspond to different forms of $\bfb(\bfX,T)$ and $\bfb_*(\bfX,T)$. In each state, every particle of an ensemble has a unique nondeterministic trajectory that satisfies the stochastic differential equation Eq.~(\ref{eq:Langevin2}). The mapping onto the Schr\"odinger equation relates $\hbar$ to the diffusion constant of the stochastic process, leading to $D=\hbar/2m$. It may seem odd that $D$ depends on the particle mass in this way, but we note that in the field theory generalization the analogy to the mass here is $m=1$ for canonically normalized fields \cite{Guerra:1973ck}. 

With one caveat (see below for Wallstrom's objection) any solution to the Schr\"odinger equation corresponds to a stochastic system with  specified drift velocities $\bfb(\bfX,T)$ and $\bfb_*(\bfX,T)$, and any drift velocities consistent with Eq.~(\ref{eq:a}) give a solution to the Schr\"odinger equation.  Simulations have verified in certain simple examples that the Schr\"odinger probabilities agree with both ensemble probabilities and with probabilities of individual particle configurations studied over time (in both configuration space and phase space) \cite{Olavo}.

We pause for the remainder of this section to discuss the status of  stochastic mechanics as an alternative to standard quantum theory. There is some resemblance of stochastic mechanics to the de Broglie-Bohm pilot wave picture of quantum mechanics \cite{deBroglie,Bohm:1951xw,Bohm:1951xx}, in that particles have definite trajectories guided in some way by a wavefunction that satisfies the Schr\"odinger equation. However, in the Bohmian picture particles that begin in an identical configuration would evolve identically, whereas the same is not true for the stochastic motion of particles in stochastic mechanics. Quantum phenomena such as the uncertainty principle, tunneling and interference, are manifested in stochastic mechanics despite the secondary role played by the wavefunction, and such phenomena have been demonstrated by simulations \cite{Nelson:1983fp,delaPena}. 
 
The Nelson-Newton equation Eq.~(\ref{eq:a}) is appropriate in the nonrelativistic limit, which we have presented for simplicity and because nonrelativistic stochastic mechanics is better established than its relativistic counterpart. Relativistic treatments of stochastic mechanics and stochastic field theory have been presented \cite{Guerra:1978kk,Guerra:1973ck,Dohrn:1985iu,Marra:1989bi,Guerra:1981ie,Morato:1995ty,Garbaczewski:1995fr,Pavon}, but the subject is not mature. 
Our principal interest is not particle mechanics but field theory, which benefits in this regard from its more democratic treatment of space and time. We summarize some important results in stochastic field theory for our purposes in the next section.

Construction of stochastic systems with  spin, and discussions of entanglement and Bell's inequality violation, appear in several places, for example Refs.~\cite{Petroni-Morato,Fritsche:2009xu,delaPena,Gaeta}.  The main point here is that the stochastic system is a 
hidden variable theory that is nonlocal because the drift velocities depend on all components of the system, and the rules of stochastic mechanics are   {\em constructed} so as to give rise to the predictions of standard quantum theory. The inherent nonlocality may be distasteful, as it was to Nelson \cite{Nelson-nonlocal}, but the nonlocality of the description appears to be as hidden experimentally in the stochastic framework as it is in ordinary quantum theory. In any case, quantum ontology theorems such as Bell's inequality appear to force some form of nonlocality on any realist framework of quantum theory, distasteful though that might be.
  
The most popular objection to stochastic mechanics is in regards to the multivalued nature of the phase of the wavefunction in states with nonvanishing  angular momentum \cite{Wallstrom}. Wallstrom pointed out that it is natural  at first to assume that $S$ defined by Eq.~(\ref{eq:v}) is a single-valued function on spacetime, but in that case not all solutions to the Schr\"odinger equation are described by a corresponding stochastic system; in particular, states with orbital angular momentum would be excluded. On the other hand, if $S$ can be multivalued, then the restriction that $S$ be allowed to change only by integral multiples of $2\pi\hbar$ around closed loops so that the wavefunction remains single-valued seems {\em ad hoc} and poorly motivated \cite{Wallstrom}. However, it was pointed out by Fritsche and Haugk that the restriction to only those multivalued phases allowed in the standard quantum theory is demanded by unitarity (conservation of probability) if we assume that solutions for $\psi$ can be superposed \cite{Fritsche:2009xu}. In other words, only those multivalued phases allowed by quantum theory are consistent in the stochastic framework if we assume the superposition principle, which seems a natural assumption once the Madelung equations have been made linear by combining them into the Schr\"odinger equation. The  author's opinion is that Wallstrom's criticism is therefore without merit. There is not a consensus opinion on this issue \cite{delaPena,Derakhshani:2015cda,Schmelzer,Bacciagaluppi}.

One additional comment deserves to be made regarding   Wallstrom's criticism in the specific context of the framework presented in this work.  Multivalued phases of the wavefunction are possible when the wavefunction has nodal manifolds. In such situations there are points at which  the phase of the wavefunction is not well defined, so the drift velocities risk obtaining singularities which would naturally be considered unphysical and wanting of some sort of regularization. Nelson made the point that stochastic trajectories are repelled by the nodal surfaces, so associated singular behavior does not lead to singular dynamics \cite{Nelson:1983fp}. However, if one is nevertheless concerned about such singularities, then it is worth noting that in the stochastic framework of this paper, the continuous stochastic system is only an approximation; at ultra-short distances smaller than the  spacing between critical points, physics is classical and particles appear free. Hence, it is reasonable to expect that singularities at nodal surfaces would be resolved due to the granularity of the stochastic impulses. The transition between the continuous stochastic mechanics and the granular stochasticity of the present framework deserves further study.

The most serious remaining concern before one can adopt stochastic mechanics as an alternative to quantum theory is a lack of understanding of the effect of a measuring apparatus on the state of the system as described by $\bfb(\bfX,T)$ and $\bfb_*(\bfX,T)$. One way or another, as the stochastic system is altered by measurement a collapse of the wavefunction of sorts takes place. The drift velocities contain epistemic information regarding the lack of knowledge of the particle trajectory.
The pressing question is how it is that the interaction with the measuring apparatus alters the drift velocities in precisely the manner to correspond to a wavefunction collapse. A better understanding of the effects of measurement in the stochastic setting would necessarily improve our understanding of this framework, including issues related to entanglement and nonlocality. It is also possible that further consideration of the issue of measurement will lead to testable distinctions between stochastic mechanics and standard quantum theory, or may even require dismissal of the stochastic framework altogether.

\section{Stochastic Field Theory and Emergent Gravity}
Relativistic quantum field theory subsumes  quantum mechanics as a more fundamental description of nature. The rules of quantum field theory give rise to a vastly more exotic variety of phenomena than quantum mechanics, including nonconservation of particle number and confinement of certain charges. It is therefore critical to the program put forward here that a stochastic formulation of quantum field theory be available. Some substantial work in this direction was presented by Guerra and Ruggiero \cite{Guerra:1973ck}, Nelson \cite{Nelson-fieldtheory,Nelson-fieldtheory2} and others. 

Guerra and Ruggiero took advantage of the representation of the free relativistic scalar field as a collection of nonrelativistic simple harmonic oscillators to define the stochastic field theory by the rules of stochastic mechanics. A state in the system is defined by a forward and backward Langevin equation for each momentum mode of the relativistic field. There is a remarkable relationship between the ground state of the relativistic quantum field and the Euclidean Markov field \cite{Guerra:1973ck}, which may ultimately prove useful for computations in the stochastic theory. In scalar theories with perturbative interactions one can  develop a perturbation theory about the ground state of the free field \cite{Guerra:1981ie}. As long as a field theory can be described in terms of interactions between modes, each of which is described quantum mechanically, then there can be expected to exist an extension of stochastic mechanics to the field theory, as in Guerra and Ruggiero's analysis of the free scalar field. However, to the author's knowledge a rigorous demonstration of the equivalence of generic nonperturbative quantum field theories, and a stochastic version of the same theories, is not available.

Consider  a free real scalar field $\phi(x)$, with action \begin{equation}
S=\int d^dx\,\frac12\left(\partial_\mu\phi\,\partial_\nu\phi\,\eta^{\mu\nu}-m^2\phi^2\right). \end{equation}
Decomposing $\phi(\mathbf{x},t)$ in momentum modes $\phi_\bfk(t)$ with $\phi_{-\bfk}=\phi_\bfk^*$, we have 
\begin{equation}
\phi(\mathbf{x},t)=\int\frac{d^{d-1}k}{(2\pi)^{d-1}} \phi_\bfk(t) e^{i\bfk\cdot\mathbf{x}},
\end{equation}
and
\begin{equation}
S=\int dt \int\frac{d^{d-1}k}{(2\pi)^{d-1}}\frac12 \left(|\dot{\phi}_\bfk|^2-(\bfk^2+m^2)|\phi_\bfk|^2\right),
\end{equation}
which describes an independent harmonic oscillator  for each $\bfk$, with  effective mass $m=1$ and frequency $\omega_\bfk=\sqrt{\bfk^2+m^2}$.  Guerra and Ruggiero's stochastic description of the free field theory is then defined by Nelson's stochastic mechanics for each momentum mode.

For example, the stochastic process for the ground state is constructed as follows. 
The Schr\"odinger-representation wavefunction formally satisfies the Schr\"odinger equation, with $\hbar$ now set to $\hbar=1$,\begin{equation}
i\frac{\partial}{\partial t} \Psi[\phi(\mathbf{x}),t]=\frac{1}{2}\int d^{d-1}x\left(-\frac{\delta^2}{\delta \phi(\mathbf{x})^2}+(\nabla \phi)^2+m^2\phi^2\right)\Psi[\phi(\mathbf{x}),t].
\end{equation}
With decomposition in momentum modes, the ground state for each mode is described by the simple harmonic oscillator ground state wavefunction\begin{equation}
\psi_\bfk(\phi_\bfk,t)=\left(\frac{\omega_\bfk}{\pi}\right)^{1/4}e^{-\omega_\bfk|\mathbf{\phi_\bfk}|^2/2}\,e^{-i\omega_\bfk t/2}.
\end{equation}
From Eq.~(\ref{eq:v}) we define, \begin{equation}
v_\bfk=2\frac{\partial S_\bfk}{\partial \phi_\bfk^*} =0,
\end{equation}
where we have identified $S_\bfk=-\omega_\bfk t/2$;
and as in Eq.~(\ref{eq:u}) we define, \begin{equation}
u_\bfk=2\frac{\partial \ln|\psi_\bfk|}{\partial \phi_\bfk^*}=-\omega_\bfk\phi_\bfk.
\end{equation}
The factors of $2$ in  $u_\bfk$ and $v_\bfk$ are due to the decomposition of the real scalar field in terms of complex momentum modes, so that $\phi_\bfk^*$ appears in the wavefunction $\Psi[\phi(\mathbf{x}),t]$ twice, from the wavefunctions of the two modes $\psi_\bfk$ and $\psi_{-\bfk}$.
Then, with $b_\bfk=v_\bfk+u_\bfk$ and $b_{*\bfk}=v_\bfk-u_\bfk$, we have the result that the ground state of the harmonic oscillator is described by the mean-reverting Ornstein-Uhlenbeck process in both the forward and backward direction, with \begin{equation}
d\phi_\bfk=-\omega_\bfk \phi_\bfk\,dt+d\xi_\bfk.
\end{equation}
Formally, we can write this in terms of $\phi(\mathbf{x})$ as,
\begin{equation}
d\phi(\mathbf{x};t)=-\sqrt{-\nabla^2+m^2}\,\phi(\mathbf{x};t)\,dt+d\xi(\mathbf{x};t).\end{equation}
The Brownian process $d\xi(\mathbf{x};t)$ contributes to the change in $\phi(\mathbf{x})$ over time $dt$. The special role of time $t$ here is appropriate for the Schr\"odinger representation, and does not imply a violation of Lorentz invariance. 

With the ground-state process for the free field at hand, it is possible to develop a perturbation theory about the free theory \cite{Guerra:1981ie}. 
Although the stochastic description of more general quantum field theories requires additional clarification,
 the field-theory generalization of the 
treatment of impulses at critical points is straightforward. Consider the diffeomorphism-invariant scalar field theory with action \cite{Carone:2016tup},
\begin{equation}
S= \int d^dx\left(\frac{\frac{d}{2}-1}{V(\phi^a)}\right)^{\frac{d}{2}-1}\sqrt{\left|\det\left(\sum_{M,N=0}^{d-1}\partial_\mu X^M\partial_\nu X^N\eta_{MN}-\sum_{a=1}^N\partial_\mu\phi^a\partial_\nu\phi^a\right)\right|},
\label{eq:Sfield}
\end{equation}
where $\eta_{MN}$ has elements equal to those of the Minkowski metric with mostly negative signature.  

\begin{figure}
\centering
\includegraphics[scale=.5]{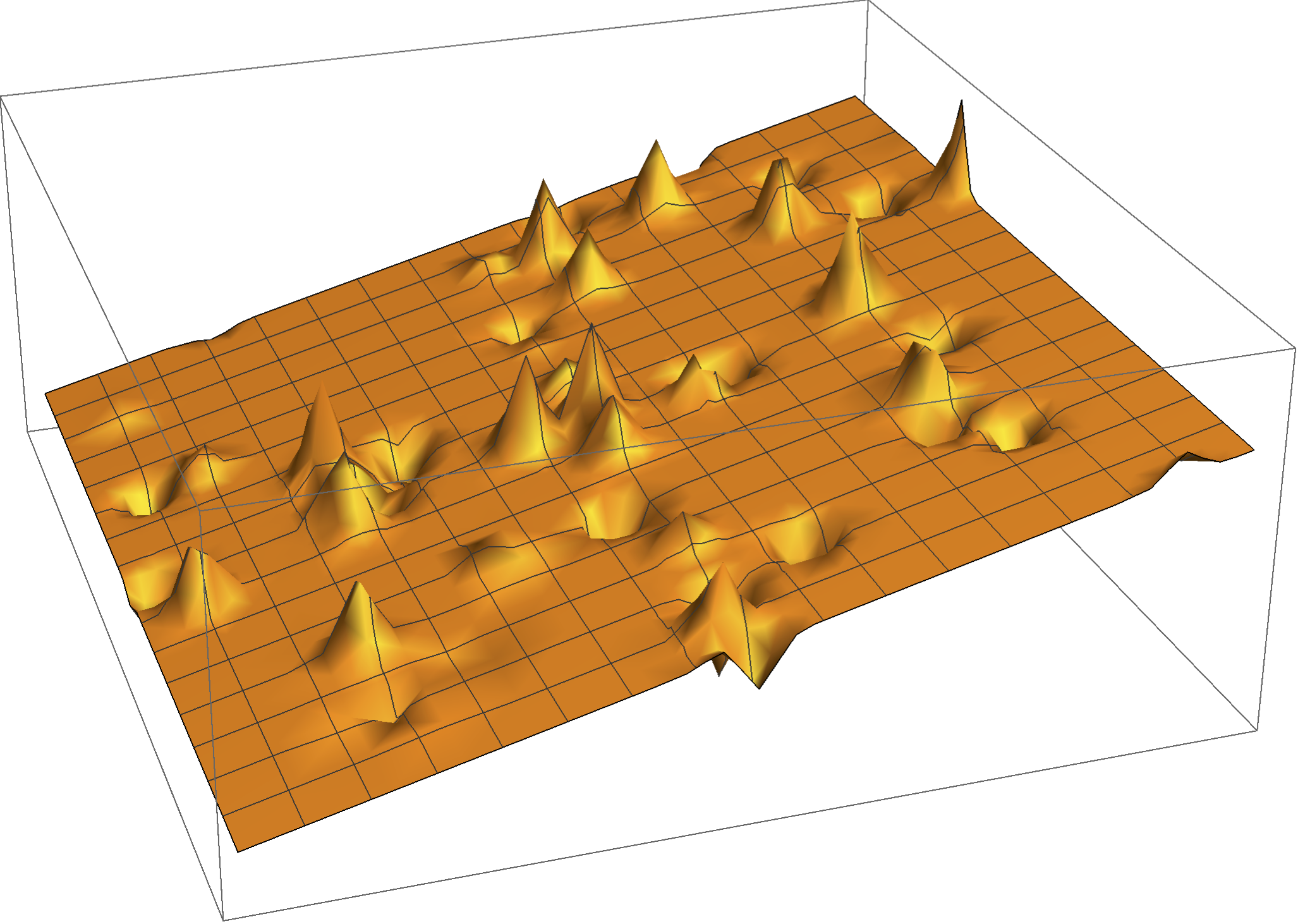}
\caption{Clock (or ruler) field as function of coordinates. The values of the clock and ruler fields at the critical points are invariant under diffeomorphisms. The shape of the hills and valleys is coordinate dependent.}
\label{fig:spacetime}
\end{figure}

The theory defined by Eq.~(\ref{eq:Sfield}) was studied in Refs.~\cite{Akama:1978pg,Akama:2013tua,Carone:2016tup,Chaurasia:2017ufl,Carone:2017mdw} from the perspective of emergent gravity. In $d$ spacetime dimensions, $d$ fields $X^M$ are identified with clock and ruler fields. The critical points are points where $\det\left(\partial X^M/\partial x^\mu\right)=0$. The existence of  critical points, and the values of the clock and ruler fields at those points, are invariant under nonsingular coordinate transformations. We assume a random profile of the clock and ruler fields, so that the underlying spacetime is topologically as in Fig.~\ref{fig:spacetime}. More complicated spacetime foams with handles could be considered, but are beyond the scope of the present discussion. There is no meaning to the geometry of the clock and ruler field configurations {\em a priori}, as the theory is invariant under coordinate transformations without reference to a spacetime metric over the coordinates $x^\mu$. A notion of spacetime geometry emerges only after quantization, at which point the operator
\begin{equation}
g_{\mu\nu}=\frac{d/2-1}{V(\phi^a)}\left(-\sum_{I,J=0}^{d-1}\partial_\mu X^M\partial_\nu X^M\eta_{MN}+\sum_{a=1}^N\partial_\mu\phi^a\,\partial_\nu\phi^a\right)\label{eq:metric}
\end{equation}
assumes the role of the spacetime metric in the effective gravitational theory at long distances \cite{Carone:2016tup}.

The equations of motion can be written formally in terms of $g_{\mu\nu}$, its inverse $g^{\mu\nu}$ and its determinant $g\equiv\det g_{\mu\nu}$, \begin{eqnarray}
\partial_\mu\left(\sqrt{|g|}g^{\mu\nu}\, \partial_\nu \phi^a\right)&=&-\sqrt{|g|}\frac{\partial V}{\partial \phi^a}, \label{eq:phiEOM} \\
\partial_\mu\left(\sqrt{|g|}g^{\mu\nu}\, \partial_\nu X^M\right)&=&0. \label{eq:relative-harmonic}
\end{eqnarray}
Eq.~(\ref{eq:phiEOM}) is the equation of motion for a scalar field in a curved background with metric $g_{\mu\nu}$, but $g_{\mu\nu}$ here is the composite operator defined in Eq.~(\ref{eq:metric}). There are solutions of the form $\phi[X^M(x)]$ for any clock and ruler field profiles $X^M(x)$. Eq.~(\ref{eq:relative-harmonic}) can be considered a harmonic gauge condition relative to the clock and rulers $X^M$. In the absence of critical points the fields $X^M$ could be transformed to the Monge gauge $X^M\propto \delta_{\ \mu}^M x^\mu$ by a diffeomorphism, in which case Eq.~(\ref{eq:relative-harmonic}) would be equivalent to the usual harmonic gauge condition, \begin{equation} 
\partial_M\left(\sqrt{|g|}g^{MN}\right)=0. \end{equation}

If there are critical points where $\det(\partial X^M/\partial x^\mu)=0$, then hypersurfaces of constant $X^M(x)$ (for each $M$) need not be pathwise connected. Solutions to the equations of motion of the form $\phi(x)=\phi[X^M(x)]$ remain consistent as long as the appropriate square root branch choices are made on each disconnected region of constant $X^M$. There is therefore no global problem of spacetime in the classical setting. Adding stochasticity to the fields, we no longer have $\phi(x)=\phi[X^M(x)]$ precisely; there can be effective jumps in the field configuration from one disconnected region of fixed $X^M$ to another. All of this is analogous to the discussion of the relativistic particle in Sections~\ref{sec:ClockRulers} and \ref{sec:StochasticMechanics}.

\begin{figure}
\centering
\includegraphics[scale=.5]{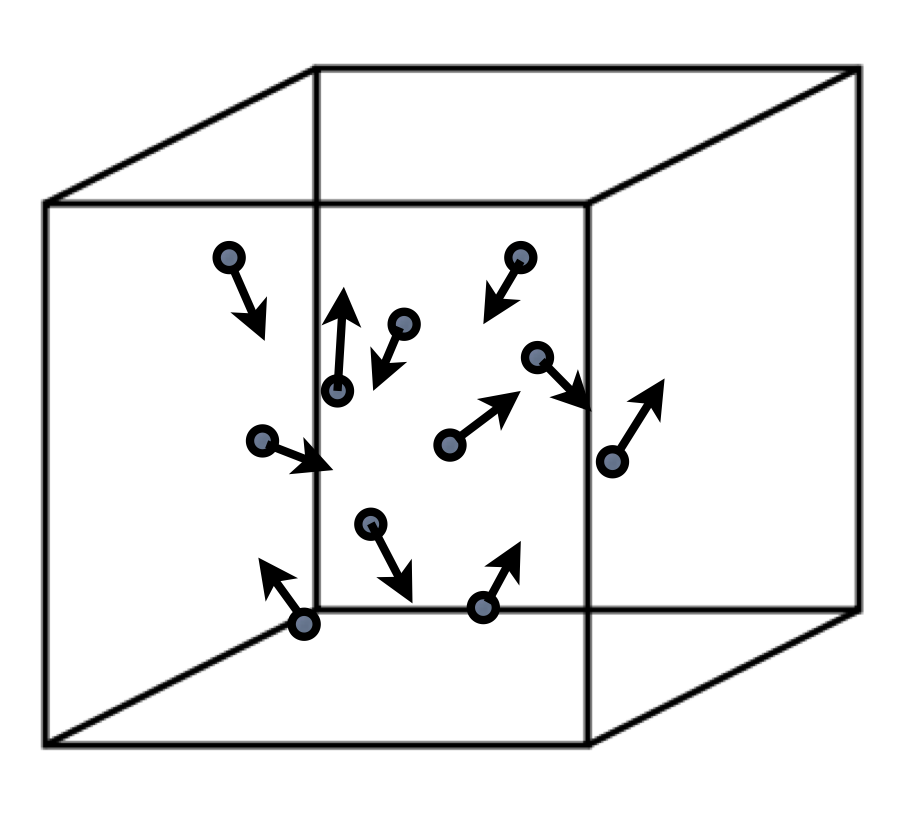}
\caption{Critical points in the space of clock and ruler fields. At each critical point there is a corresponding critical direction, denoted by an arrow, in which the physical fields effectively jump due to the stochastic impulse in that direction.}
\label{fig:critical}
\end{figure}

We assume that the stochastic impulses to the physical fields take place only at the critical points, which is not fundamentally necessary but simplifies the discussion. Then, at each critical point we identify a critical direction with respect to which we associate  effective jumps of the physical fields, in analogy with the Langevin-like equation Eq.~(\ref{eq:Langevin}). This can be visualized in terms of the space of critical points supplemented with an element of the tangent space  $\mathbb{R}^d$ that indicates the direction of impulses for the remaining fields, as in Fig.~\ref{fig:critical}.
The $d\times d$ matrix $\partial X^M/\partial x^\mu$ is singular at each critical point, and the left zero-eigenvector $V_M$, with $V_M\,\partial X^M/\partial x^\mu=0$, defines the corresponding critical direction. Therefore,
\begin{equation}
\Pi^a_{\rm crit}\equiv V^M \,\frac{\partial \phi^a[X]}{\partial X^M} 
\end{equation}
is assumed to change stochastically following some probability distribution at the corresponding critical point. The distribution of impulses gives rise to a Langevin description of this system in the spirit of Guerra and Ruggiero's formulation of the stochastic quantum field.  
Just as we were able to define an effective forward-in-time trajectory with stochastic jumps in the stochastic mechanics developed in the previous section, we define an effective description in which $X^M$ may be put into Monge gauge, $X^M=c\,\delta^M_{\ \mu}\,x^\mu$ for some conveniently chosen constant $c$, effectively excising the critical points from the smooth  spacetime of the emergent quantum theory. The effective jumps in the field are incorporated into the Langevin description. 
We assume that fields can be decomposed in momentum modes, each of which is described quantum mechanically as in Guerra and Ruggiero's stochastic free theory. At least in a perturbative setting the stochastic system appears to be well defined \cite{Guerra:1981ie}.

The equivalence of the stochastic theory defined in this way and ordinary quantum field theory deserves examination by way of simulations. However, if we grant equivalence of the stochastic theory at long distances with the related quantum field theory, the stochastic construction includes a regulator scale related to the distribution of critical points. Between critical points, physics is classical in this framework. With a covariant regulator in hand, we then have the additional result that quantum gravity should arise as an emergent interaction at long distances \cite{Sakharov:1967pk}.  

With dimensional regularization as a proxy for a physical covariant regulator, scattering amplitudes in these models have been calculated in an expansion about a flat background defined by Eq.~(\ref{eq:metric}) with $\phi^a=0$ and $X^M=c\,\delta^M_{\ \mu}\,x^\mu$, as in Refs.~\cite{Carone:2016tup,Carone:2017mdw}. An expansion of the action about the flat background includes both the Lorentz-invariant scalar field action with potential $V(\phi)$, and to next order a quartic interactions that depend on the energy-momentum tensor of the physical fields. The scattering amplitudes of the physical fields feature a long-range gravitational interaction coupled to the matter energy-momentum tensor with a coupling suppressed by a scale explicitly related to the field theory regulator. At leading order in the gravitational coupling two-into-two scattering was shown to agree with general relativity. Certain gravitational self-interactions were also shown to agree with the predictions of general relativity. Here we call such models induced emergent gravity models to distinguish them from other varieties of emergent gravity scenarios such as those based on the AdS/CFT correspondence \cite{Maldacena:1997re}, entropic gravity \cite{Verlinde:2010hp} and ideas suggesting entanglement as an origin of spacetime \cite{VanRaamsdonk:2010pw,Maldacena:2013xja,Cao:2016mst}; and also to distinguish these models from Sakharov's original discussions of induced gravity in which gravitation was semiclassical rather than quantum \cite{Sakharov:1967pk}. 

It is important to note that the present construction is background independent, despite its analysis by a perturbative expansion about Minkowski space. No spacetime metric explicitly appears in Eq.~(\ref{eq:Sfield}), and the existence (or not) of a Lorentz-invariant state about which gravitation can be studied perturbatively is a consequence of the dynamics of the theory. Two things are crucial for gravity to emerge in precisely this way: the stochastic theory must be diffeomorphism invariant with a diffeomorphism-invariant regulator; and the states with critical points of the clock and ruler fields are excised from the functional integrals in the quantum field theory because of the manner in which the distribution of critical points is related to quantization in the first place. 

Finally, we consider the stochastic theory beginning with classical action Eq.~(\ref{eq:Sfield}) in terms of the constructive procedure for stochastic emergent quantum gravity outlined in the introduction. 
\begin{enumerate}
\item We begin with the theory of $N$ real scalar fields with potential $V(\phi^a)$, with action,
\begin{equation}
S_0=\int d^dx\left(\frac{1}{2} \sum_{a=1}^N\partial_\mu\phi^a \partial_\nu\phi^a \eta^{\mu\nu}-V(\phi^a)\right). \end{equation}
We add to the theory $d$ massless scalar clock and ruler fields $X^M$, so that the action becomes,
\begin{equation}
S_1=\int d^dx\left(-\frac{1}{2} \sum_{M,N=0}^{d-1}\partial_\mu X^M \partial_\nu X^N \eta_{MN}\,\eta^{\mu\nu}+\frac{1}{2}\sum_{a=1}^N\partial_\mu\phi^a \partial_\nu\phi^a \eta^{\mu\nu}-V(\phi^a)\right). \end{equation}
The clock field $X^0(x)$ has the wrong-sign kinetic term, but the field will be gauged away.
\item
To couple the theory to an auxiliary spacetime with metric $g_{\mu\nu}$, we simply replace $\eta_{\mu\nu}$ in $S_1$ with $g_{\mu\nu}$, giving us
\begin{equation}
S_2=\int d^dx\left(-\frac{1}{2} \sum_{M,N=0}^{d-1}\partial_\mu X^M \partial_\nu X^N \eta_{MN}\,g^{\mu\nu}+\frac{1}{2}\sum_{a=1}^N\partial_\mu\phi^a \partial_\nu\phi^a g^{\mu\nu}-V(\phi^a)\right). \label{eq:S2} \end{equation}
In a more general theory we would introduce the vielbein $e_\mu^m$ and couple the fields covariantly to the vielbein. 
\item
The constraint $T_{\mu\nu}[g_{\alpha\beta},X^M,\phi^a]=0$ has Eq.~(\ref{eq:metric}) as solution for $g_{\mu\nu}$. In a more general theory, the vielbein is determined by the constraint of vanishing $T_{\mu\nu}$ up to local Lorentz transformations. However, the constraint cannot generally be solved in closed form.
In the present toy model, substitution of Eq.~(\ref{eq:metric}) into Eq.~(\ref{eq:S2}) gives the Nambu-Goto-like action Eq.~(\ref{eq:Sfield}).
\item
Assume a random profile for the clock and ruler fields, with typical spacing between critical points comparable to the Planck scale.
Construction of the stochastic quantum theory proceeds as in the earlier discussion of this section. Stochastic impulses for bosonic fields are imparted at the critical points of the clock and ruler fields as described. 

For fermionic fields the interpretation of the impulses is more subtle, but we can formally write a Langevin equation for  fermionic fields, where the stochastic differentials and drifts are interpreted as Grassmann variables \cite{Damgaard:1983tq}.   A functional-integral representation of the Schr\"odinger wavefunction  generalizing the Feynman-Kac formula may be available. To the author's knowledge a procedure like this for fermions has not been carried out in detail.
\end{enumerate}

Our conjecture is that quite generally the stochastic theory constructed from a classical field theory as outlined above is equivalent to standard quantum field theory with emergent gravity at distances much longer than the typical separation between critical points of the clock and ruler fields.  In practice, calculations can be done with any covariant regulator. In the emergent quantum field theory, the clock and ruler fields can be fixed to Monge gauge $X^M\propto x^\mu\delta^M_\mu$.

\section{Discussion}
We have provided evidence that stochastic evolution of otherwise classical relativistic fields can at the same time be responsible for quantization and for gravitational interactions described by general relativity at long distances.  Key to the construction is the incorporation of scalar fields that play the role of clock and rulers relative to which the dynamics is defined. Global Lorentz transformations act on the scalar clock and ruler fields, while diffeomorphisms  act as a reparametrization of all the fields in the theory. With the clock and ruler fields included in the action following the prescription described in the text, the low-energy, weak-field effective description includes the Lorentz-invariant theory that begins the construction.
We assume that there is a nonzero scale associated with the typical distance between stochastic impulses, so  physics at ultra-short distances  is classical, and spacetime is continuous. The unusual feature of spacetime in this framework is that at the scale where the stochasticity is evident, hypersurfaces of fixed clock time or ruler position  need not be pathwise connected. We have explained that such behavior fits naturally into the framework of Nelson-Guerra-Ruggiero's stochastic quantum field theory.

There are many loose ends to tie before  stochastic emergent quantum gravity can be considered a complete description of nature. Among these is the need for an understanding of the effects of measurement on the state of the system as described by drift velocities. While the rules of standard quantum theory including wavefunction collapse and entanglement may be incorporated by construction, as a prequantum description it seems possible to address the measurement problem concretely in the stochastic framework. The stochastic description of Grassmann fields also requires further study in order to determine whether the outlined stochastic construction in fact reproduces ordinary quantum field theory with fermions as conjectured. There is also the question of whether the presumed stochastic impulses are irreducible or are due to interactions with some sort of background. A proposal of the latter type is stochastic electrodynamics, and is discussed at length in Ref.~\cite{delaPena}. 

A quantum theory of gravity should be expected to resolve the modern puzzles of gravitational physics. The prevailing wisdom today is that holographic aspects of gravity should guide our thinking about quantum gravity  \cite{tHooft:1993dmi,Susskind:1994vu,Maldacena:1997re}, but such aspects would appear to be secondary in the stochastic construction. What is the face of holographic aspects of quantum gravity in the stochastic framework?  
Do black-hole horizons violate the equivalence principle by creating a firewall \cite{Almheiri:2012rt}? What becomes of black-hole and cosmological singularities?  Why does our universe appear to have such an unnaturally small cosmological constant responsible for the accelerated expansion of the universe? 
  
Black-hole geometries are nonperturbative in the present approach, and the resolutions to puzzles regarding black hole states in this framework are presently unclear. However, one can hope that simulations of the stochastic field theory will provide answers to  questions related to black holes and holography.  Simulations that would probe emergent gravitational effects would need to be sensitive to long-range interactions that decouple in the continuous stochastic limit. However, one thing that we can say already is that in the framework presented here diffeomorphism invariance is necessary as a fundamental principle in order for the clock and ruler fields to play the role that they do, and quantum theory is modified at short distances. Hence, one might guess that the resolution to the firewall puzzle will lie in the modification of quantum theory and not in a violation of the equivalence principle. In Monge gauge the field-space metric of the clock and ruler fields becomes a background spacetime metric \cite{Chaurasia:2017ufl}, so it might also be fruitful to probe aspects of holography by replacing the flat field-space metric of our construction with a black-hole or Anti-de Sitter metric.

Despite the global Lorentz invariance that acts on the clock and ruler fields in the action Eq.~(\ref{eq:Sfield}), the Lorentz symmetry does not ensure stability of Lorentz-invariant states.
The cosmological constant has to be tuned in order for the Lorentz-invariant vacuum to be perturbatively stable in induced emergent gravity models \cite{Sakharov:1967pk,Akama:1978pg,Carone:2016tup}.  As in any theory of gravity that has not been well studied, we may hope that the theory somehow hides any large cosmological constant by effects that are not apparent at lowest order in perturbation theory. For now the tuning of the cosmological constant appears to be required in order to match observational data. 

\section{Conclusions}
It is possible that quantum field theory breaks down as a description of nature at distances and times shorter than some fundamental scale.  In that case, it is natural to expect the more primitive physical framework to regularize ultraviolet divergences in the quantum field theory that emerges as a description of physics at longer distances. Assuming diffeomorphism invariance in the fundamental description, gravitation will generally arise as an emergent interaction, with gravitational coupling related to the fundamental short-distance scale at which the quantum field theory emerges.
If the stochastic version of the story told here survives scrutiny as an origin for quantization and gravitation, then quite remarkably both the existence of atoms and the motion of planets would be consequences of random fluctuations  associated with the folding of fundamental rulers and the jittery nature of clocks.

\begin{acknowledgments}
This work was supported by the NSF under Grant PHY-1519644. The author is grateful to Christopher Carone, Diana Vaman, Shikha Chaurasia and Yiyu Zhou for many enlightening conversations and collaboration on emergent gravity models, to Eddy Chen and Daniel Sudarsky for stimulating conversations regarding foundations of quantum theory, and to Christopher Carone for comments on the manuscript.
\end{acknowledgments}


\end{document}